\documentclass[fleqn]{an}
\usepackage{graphicx}
\usepackage{times}
\begin{document}

\Pagespan{1004}{}

\Yearpublication{2008}%
\Yearsubmission{2008}%
\Month{9}%
\Volume{329}%
\Issue{9/10}%
\DOI{10.1002/asna.200811056}
\title{Massive Black Hole recoil in high resolution hosts}

\author{J. Guedes\inst{1}\fnmsep\thanks{Corresponding author: {javiera@ucolick.org}}, J. Diemand\inst{1}, M. Zemp\inst{1}, M. Kuhlen\inst{2}, P. Madau\inst{1}, L. Mayer\inst{3,4}, \and  J. Stadel\inst{3}
}
\titlerunning{Massive Black Hole recoil in high resolution hosts}
\authorrunning{J. Guedes et al.}
\institute{Department of Astronomy and Astrophysics, University of California,
1156 High Street, Santa Cruz, CA 95062, USA 
\and 
Institute for Advanced Study, Einstein Drive, Princeton, New Jersey 08540, USA
\and
Institute for Theoretical Physics, University of Zurich, Winterthurer Strasse190, CH-8057 Zurich, Switzerland 
\and 
Institut fur Astronomie, ETH Zurich, Wolfgang-Pauli-Strasse 16, CH-8093 Zurich, Switzerland 
}

\received{2008 Sep 18}
\accepted{2008 Sep 19}
\publonline{2008}


\abstract{%
The final inspiral and coalescence of a black hole binary can produce highly
beamed gravitational wave radiation. To conserve linear momentum, the black hole
remnant can recoil with ``kick" velocity $v_\mathrm{kick}$ $\leq$ 4000 km/s. We present two sets of full N-body simulations of recoiling massive black holes (MBH) in high-resolution, non-axisymmetric potentials. The host to the first set of simulations is the main halo of the Via Lactea I simulation (Diemand et al. 2007). The nature of the resulting orbits is investigated through a numerical model where orbits are integrated assuming an evolving, triaxial NFW potential, and dynamical friction is calculated directly from the velocity dispersion along the major axes of the main halo of Via Lactea I. By comparing the triaxial case to a spherical model, we find that the wandering time spent by the MBH is significantly increased due to the asphericity of the halo. For kicks larger than 200 km/s, the remnant MBH does not return to the inner 200 pc within 1 Gyr, a timescale an order of magnitude larger than the upper limit of the estimated QSO lifetime. The second set of simulations is run using the outcome of a high-resolution gas-rich merger (Mayer et al. 2007) as host potential. In this case, a recoil velocity of 500 km/s cannot remove the MBH from the nuclear region.}
\maketitle

\section{Introduction}

In the context of the currently favored  $\Lambda$CDM cosmogony, where large halos are assembled through the hierarchical merging and accretion of small progenitors, MBH mergers should be common. While dynamical friction against the dark matter and baryons can only 
form binaries with separations of $\sim$1 pc, other dynamical process such as 
stellar ejection and gas drag can efficiently reduce the binary orbital separation 
down to $\sim$0.001 pc, the regime where gravitational wave radiation dominates 
the orbital energy loss.
This radiation is typically anisotropic due to asymmetries associated with the black holes' 
mass and spin, causing the center of mass of the system to recoil in order 
to balance the linear momentum carried away by the gravitational wave radiation 
(\cite{bekenstein73}; \cite{fitchett84}; \cite{favata04}).  The ``kick'' velocity of the remnant depends on the mass ratio ($m_2/m_1 < 1$) and the spin parameters ($a_1,\, a_2$) of 
the binary, but not on the total mass of the system.  Early estimates of the 
recoil velocity, framed in the post-Newtonian regime for non-spinning, 
unequal mass black holes (\cite{fitchett83}; \cite{redmount89}), yielded velocities 
in the range $100< v_\mathrm{kick}< 500$ km/s which were
successfully reproduced by numerical data (\cite{baker06b}; \cite{gonzalez07b};  
\cite{herrmann07}).  Simulations with varying (arbitrary) spin orientations 
(\cite{campanelli07b}) and mass ratios (\cite{baker08}) show that recoil 
velocities can be significantly larger, reaching ${v_\mathrm{kick} \sim 2000}$ 
km/s (\cite{campanelli07a}; \cite{gonzalez07b}) and are predicted to be as 
large as $v_\mathrm{kick} \sim 4000$ km/s for unequal mass ($m_2/m_1 = 1/3$), 
maximally spinning black holes (\cite{campanelli07a}). In order to address the issue 
of whether off-nuclear QSO/AGN are common and 
detectable, we ought to understand the dynamics of their orbits. The radial 
orbit of a recoiling MBH in a spherically symmetric potential  was studied 
analytically by \cite{madau04}. They showed that large kicks (400 km/s) can 
displace MBHs a few tens of kiloparsecs away from the center of a Milky-Way 
size stellar bulge. After the kick, the MBH undergoes several oscillations 
before decaying back to the bottom of the potential. Most of the orbital 
energy is lost during the MBH passages through the center, where dynamical
friction is most efficient. \cite{gualandris07} substantiated these results 
by performing direct summation N-body simulations of MBH recoil in a 
spherically symmetric galaxy, modeled as a binary-depleted core-S\'{e}rsic 
profile (\cite{graham03}). They find that after reaching the core radius, the MBH and the core experiences damped oscillations with bulge's center of mass, which decelerate the remnant MBH until it reaches thermal equilibrium with the surrounding 
stars.  They estimate that this final stage may take up to 1 Gyr in large 
galaxies. As we show in Sect.~\ref{tri_model}, the MBH wandering time can be largely increased if the host potential is triaxial, while the presence of gas can significantly damp the overall MBH motion. 

After a major merger, however, the remnant dark matter halo tends to be prolate (e.g. \cite{novak06}), a factor that can play a major role for large kicks. 

In the following we carry out N-body simulations of recoiling MBH in two high-resolution non-axisymmetric potentials: the Via Lactea I simulation's main halo, a triaxial dark-matter-only host with known triaxiality (\cite{kuhlen07}) and evolution parameters (\cite{diemand07}), and the gas-rich galaxy merger remnant described in \cite{mayer07}. In the dark-matter only case, we present a numerical model that successfully reproduces the main features of the MBH orbits, and can therefore be used to estimate the fallback time and maximum displacement for any given kick velocity (\cite{guedes08}, in preparation).

\begin{figure}[t]
  \hskip-3mm
\scalebox{1.06}{
\includegraphics[scale=0.45,clip]{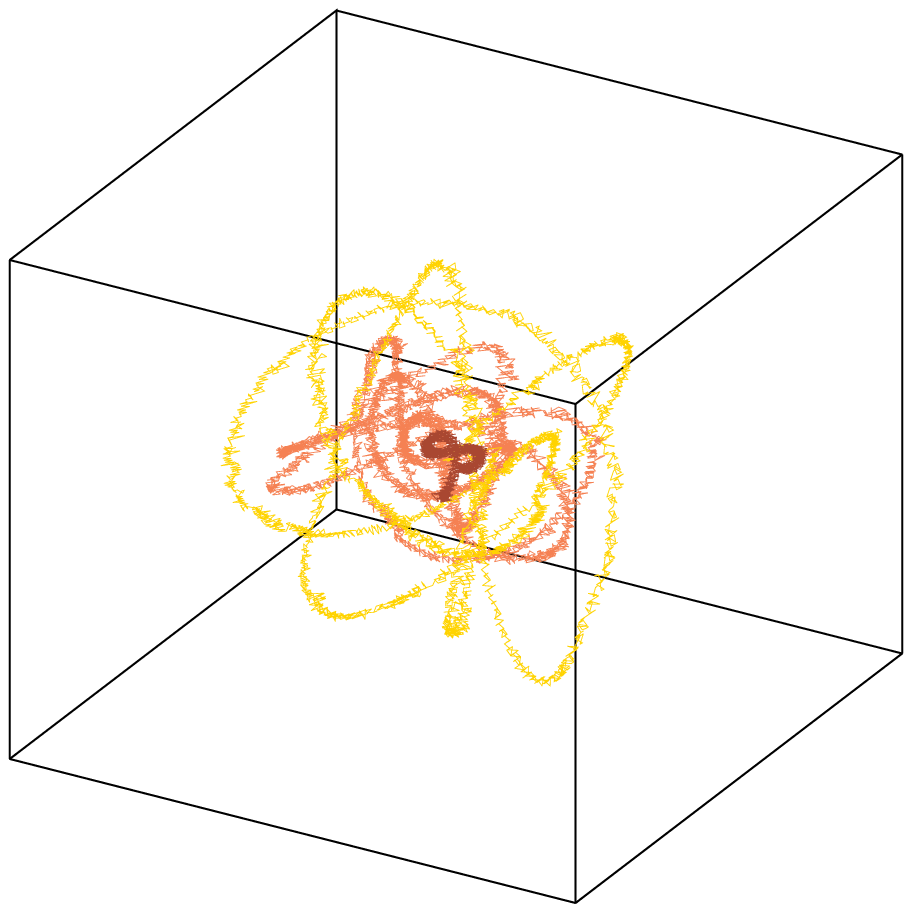}
             }
\vskip-4mm
\caption{(online colour at: www.an-journal.org) Resulting 3D orbit corresponding to simulation VL080. The first 0.5 Gyr are plotted in yellow, the following 0.5 Gyr in orange and the remaining 0.1 Gyr is plotted in dark red. The box has side length $L=1.2$ kpc.}
\label{orb3dvl01}
\end{figure}

\section{Recoil in a triaxial dark matter halo}
\subsection{Simulations}\label{dm_simulations}
Simulations of recoil in a triaxial dark matter halo are run using the entire volume of the Via Lactea I simulation (hereafter VLI), a periodic box of size $L = 90$ Mpc. The high-resolution region contains over 230 million particles of mass of ${m_{\rm p} = 2.1\!\times\! 10^4\, {\rm M}_{\odot}}$.  The exquisite resolution of VLI allows us to adopt the 
mass of SgrA* ${M_{\bullet} = 3.7 \times\! 10^6\, {\rm M}_{\odot}}$ (\cite{ghez05}) for the 
MBH and a force softening length $\epsilon$ = 90 pc. The five halo $+$ MBH 
systems are evolved for 1.1 Gyr using PKDGRAV (\cite{stadel01}), a tree 
algorithm that includes up to hexa-decapole moments to reach high accuracy 
in the force calculation. The five runs correspond to five kick velocities 
$v_\mathrm{kick}$ = 80, 120, 200, 300, and 400 km/s for runs labeled VL080 to VL400 
respectively. 

\begin{figure}[t]
\vskip-3mm \hskip-4mm
\includegraphics[scale=0.49]{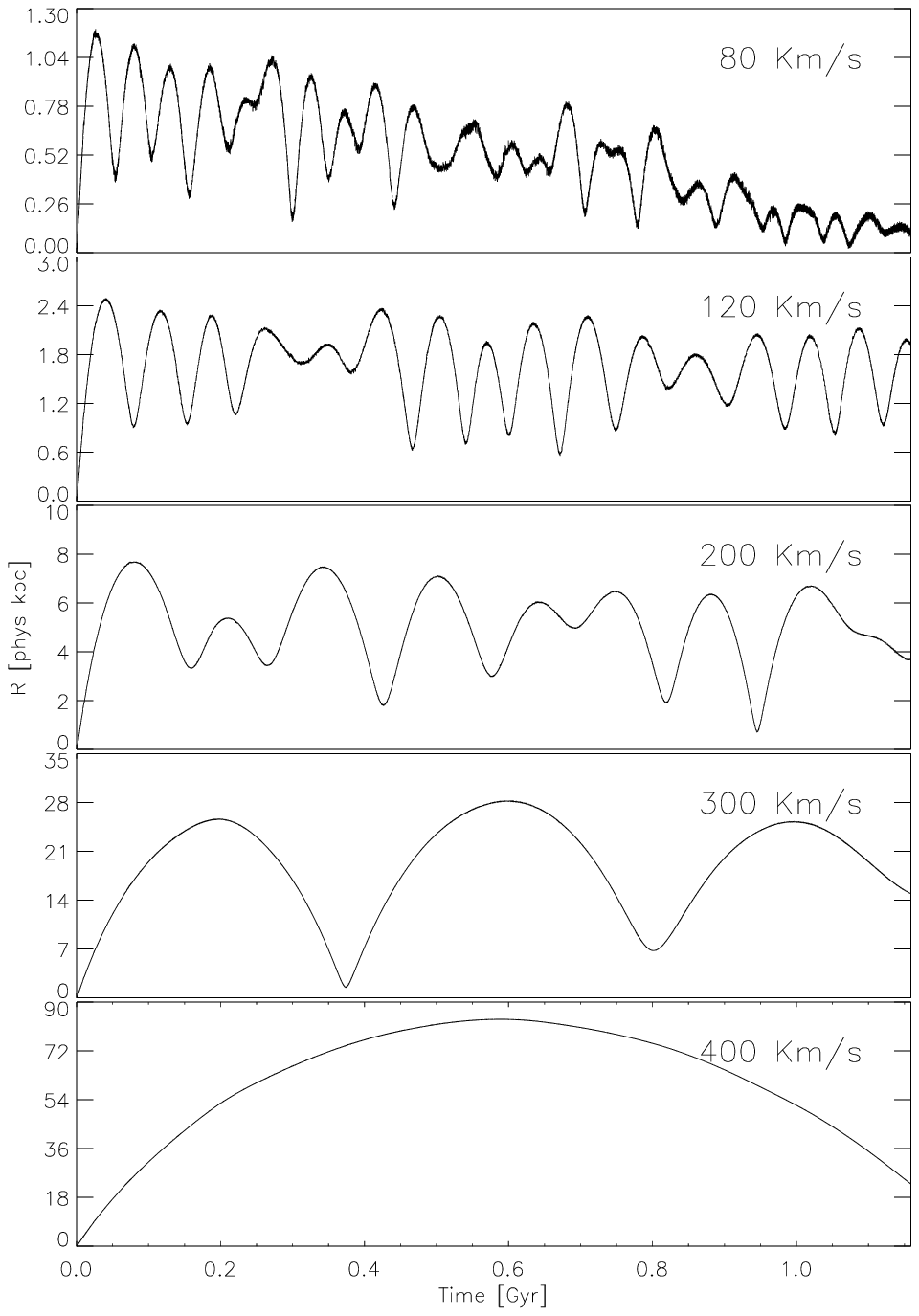}
\vskip-2mm
\caption{Simulations VL080--VL400. Resulting orbits of recoiling MBHs from $N$-body simulations in the VLI main halo.}
\label{all_runs}
\end{figure}

We place MBH at the densest point of the halo, with phase-space coordinate ${\vec w}_{\rm c}$, at an initial redshift $z_{\rm i} = 1.54$, 300 Myr after the halo suffered its last major merger. At this epoch the halo is characterized by $R_{200} = 187$ kpc and ${M_{200} = 1.02 \times\! 10^{12}\, \rm M_{\odot}}$, where $R_{200}$ is defined as the radius within which the enclosed average density is 200 times the critical density. The MBH particle is tracked at every time-step and its position and velocity are measured with respect to ${\vec w_{\rm c}}$. 

A 3D rendition of the orbit for simulation VL080 is shown in
Fig.~\ref{orb3dvl01}. Orbits for simulations VL080--VL400 are shown in
Fig.~\ref{all_runs}. These orbits sample a large volume of VLI and reveal the
asphericity of the main halo which causes angular 
momentum transfer to the $y$ and $z$ components of the MBH velocity. As will be shown 
in more detail in Guedes et al. (2008, in prep.), 
this increases the MBH wandering time, especially in cases where the kick is large enough to leave the baryonic component behind.
For simulation VL080, which features the smallest kick velocity $v_\mathrm{kick}$ = 
80 km/s, we are able to obtain both a maximum displacement of 1.2 kpc and a 
fall back time of 1.1 Gyr and therefore it can be used to constrain the semi-analytical 
model described on Sect.~\ref{tri_model}. 

\phantom{ }

\subsection{Triaxial model}\label{tri_model}

The orbit of a MBH can be characterized by a conservative force due to 
the dark matter potential $\nabla\Phi$, and a damping dynamical 
friction term. This generates a system of six coupled differential 
equations that can be separated as follows:
\begin{eqnarray}
\lefteqn{\dot{{\vec r}} = {\vec v},} \\
\nonumber 
\lefteqn{\dot{{\vec v}} = -\nabla\Phi + {\vec f}_\mathrm{DF}.}
\end{eqnarray}
The triaxial dark matter potential is modeled as modified NFW profile (\cite{navarro97}),
\begin{eqnarray}
\lefteqn{\Phi_\mathrm{NFW}^\mathrm{T} = -\frac{G M_{200}}{r_{\rm s}
f(c)}\frac{\ln(1+x_{\rm T})}{x_{\rm T}},} \\
\lefteqn{r_\mathrm{T}  =  \left ( x^2 + \frac{y^2}{p^2} + \frac{z^2}{q^2} 
\right )^{1/2},}
\end{eqnarray}
where $x_{\rm T} = r_{\rm T}/r_{\rm s}$, with $r_{\rm s}$ the scale radius, and $p$ and $q$ are the triaxiality parameters.

The classical Chandrasekhar dynamical friction formula is not valid in a triaxial system, because the  velocity dispersion is non-isotropic and the velocity distribution deviates from Maxwellian. We adopt the \cite{pesce92} generalization of the dynamical friction formula for non-axisymmetric systems, 
\begin{equation}
{\vec f}_\mathrm{DF} = -\Gamma_1 v_1 \hat{e}_1  -\Gamma_2 v_2 \hat{e}_2  
-\Gamma_3 v_3 \hat{e}_3,
\end{equation}
\noindent where $v_i$ is the component of the black hole velocity 
along the principal axis $\hat{e}_i$ of the halo's velocity dispersion tensor, 
and $\Gamma_i$ are the dynamical friction coefficients given by (\cite{pesce92};\cite{vicari07})
{\setlength{\mathindent}{0pt}
\begin{eqnarray}
\Gamma_i &=& \frac{2\sqrt{2\pi}G^2 \rho({\vec r},z)  \ln \Lambda 
(M_\mathrm{BH}+m_{\rm p})}{\sigma_1^3} \times B_i({\vec v},{\bf \sigma}),\label{tdf} \\
B_i &=& \int_0^{\infty} \frac{\exp\left(\sum_{k=1}^3 
-\frac{v_k^2/ 2\sigma_k^2}{\epsilon_k^2 + u}\right)}{\sqrt{(\epsilon_1^2 +u)
(\epsilon_2^2+u)(\epsilon_3^2+u)}}\, \frac{1}{\epsilon_i^2+u} \,{\rm d}u,
\end{eqnarray}}
with $\epsilon_i = \sigma_i/\sigma_1$ and $\sigma_1$ is the largest eigenvalue.

The six components of the symmetric velocity dispersion tensor, defined as $\sigma_{ij}^2 \equiv \langle v_i v_j\rangle - \langle v_i\rangle \langle v_j \rangle$ 
are measured as a function of radius for every output of the VLI simulation. To get the 
three principal velocity dispersions, we diagonalize $\sigma_{ij}^2$, obtaining 
the principal eigenvalues  $\sigma_1^2$, $\sigma_2^2$, and $\sigma_3^2$ as 
a function of radius. To simplify our calculations, 
we perform analytical fits to $\sigma^2_k$, 
\begin{equation}
\sigma^2_k(r,z) = A_k \times \frac{1 + D_k r^{m_k}}{1+B_kr^{n_k}}\, {\rm e}^{-r/C_k},
\end{equation}
where the best fit parameters $A, B, C, D, n, m$ vary as a function of redshift. 

\begin{figure}
\includegraphics[scale=.47]{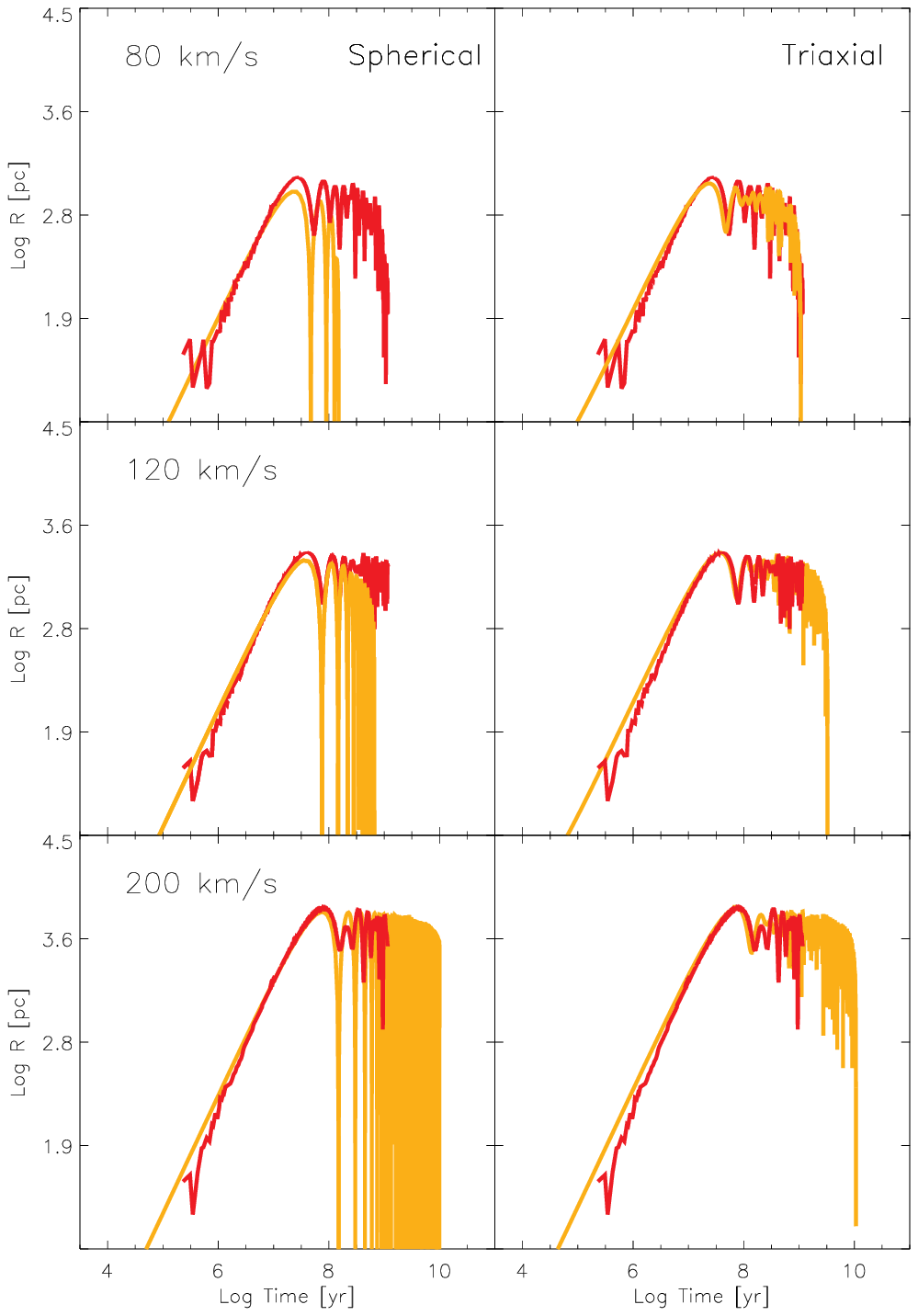}
\caption{(online colour at: www.an-journal.org) Comparison of spherical and triaxial models of the MBH orbits after suffering a gravitational wave recoil. The N-body simulation results (red) are superposed to the numerical model results (orange).}
\label{orbits_both}
\end{figure}

Several studies have shown that a constant value for the Coulomb logarithm, $\ln\Lambda$, does not match simulations results. We adopt the \cite{maoz93} formalism
\begin{equation} 
\ln \Lambda \rightarrow  \int_{r_\mathrm{min}}^{R_\mathrm{max}}\frac{\rho(r)}{\rho_0 r}\,{\rm d}r,
\end{equation}
where $\rho_0$ is the interior density, $R_\mathrm{max} = r_{\rm s}$, and we solve for $r_\mathrm{min}=G M_{\bullet}/\sigma^2(r)$, the sphere of influence of the MBH.
The resulting orbits are shown in Fig.~\ref{orbits_both} (right). Contrary to the 
spherical model (left), the orbits in a triaxial halo are non-radial, which keeps the 
MBH orbiting away from the center of the host, reducing the mean contribution of 
dynamical friction to the motion and extending the wandering time. This model 
qualitatively reproduces the results of the N-body simulations described in 
Sect.~\ref{dm_simulations}. 

\section{Recoiling MBHs in a gas merger}

We present the preliminary results of a simulation of a recoiling MBH of mass ${M_{\bullet}=3.6\times10^7\,\rm M_{\odot}}$ in a multi-component galaxy using the SPH\,+\,N-body code GASOLINE (\cite{wadsley04}). The host is the merger remnant of a high-resolution simulation by \cite{mayer07}, which features a gas particle mass of $m_{\rm p}=3000$ $M_{\odot}$, a gravitational softening $\epsilon=2$ pc, and a central MBH in each of the progenitor galaxies. The nuclear region of the merger remnant contains 2 million particles, contains a mass of ${M_{\rm g} = 3\times\! 10^9\,\rm M_{\odot}}$ and a density that ranges from $\rho = 10^{-2}$--$10^5$ atoms/cm$^3$. By the end of their simulation, the MBHs have formed a binary with semi-major axis equal to the softening length of the simulation. 

We take the simulation from this point forward. We remove both MBHs and evolve the simulation until the nucleus becomes smooth, place our MBH at the center of mass of the system, and kick it with recoil velocity $v_\mathrm{kick}=500$ km/s. In this case, the MBH was displaced only 30 pc from the center of mass. A simple Bondi accretion estimate yields an accretion rate of $M_\mathrm{acc} = 0.06 \,\rm M_{\odot}/yr$, and therefore the MBH can become an active AGN during its wandering time. 

\begin{figure}[t]
\includegraphics[scale=0.57]{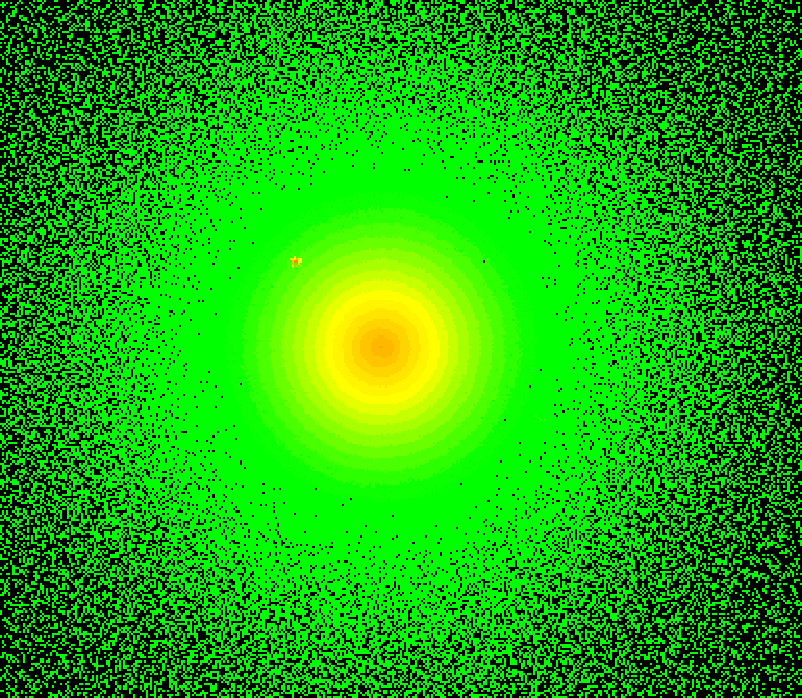}
\caption{(online colour at: www.an-journal.org) Projected density plot of the
nuclear region of the merger remnant. A wandering MBH is seen at apocenter as an
over-density near the nuclear region at ${t=5.4}$ Myr after suffering a 500 km/s
recoil kick. The image corresponds to an area of 30 pc $\times $30 pc.}\label{merger}
\end{figure}

As seen in Fig.~\ref{merger} the MBH produces an over-density signature on the
nuclear gas at the apocenter of its orbit, a feature that can be associated with
the emission of X-ray radiation (e.g. \cite{yutawa08}). 

\section{Discussion}
If MBH mergers are common in the context of $\Lambda$CDM, then the ``rocket effect" should also be common, especially at high-redshift where the bulk of the mass assembly occurs. 
Because recoiling MBHs carry gas and stars with them, the detection of 
off-nuclear QSOs is possible by finding velocity shifts between the narrow 
line region associated with gas accreting onto the MBH, and the broad line 
emission associated with the galaxy left behind (\cite{bonning07}). The best 
off-nuclear QSO candidate today is thought to be powered by a recoiling SMBH 
of mass $M_{\bullet} =6 \times\! 10^8 \,\rm M_{\odot}$ which travels at a velocity 
of $2650$ km/s with respect to the host galaxy (\cite{komossa08}). 

In order to assess the detectability of so-called naked QSOs, we ought to understand the dynamics of the  recoiling MBH orbits and the environment provided by the host potential. While detectability requires the presence of an accretion disk around the remnant MBH, the medium of its host galaxy can damp its motion significantly, disfavoring its detection. However, if the kick is large enough that, while keeping an accretion disk and stellar sphere on influence, the MBH can reach as far as the (generally) prolate dark matter halo, dynamical friction is much less efficient and therefore the detection probability is mostly determined by the lifetime of the QSO. 

\acknowledgements
This research was conducted at the Pleiades Super Computer Cluster at UCSC and
was funded by an NSF Graduate Student Research Fellowship to J. G.

\end{document}